\pdfoutput=1

\documentclass[11pt]{article}

\usepackage[final]{acl}

\usepackage{times}
\usepackage{latexsym}
\usepackage{hyperref}
\newcommand{\gray}[1]{\textcolor{gray}{#1}}
\usepackage[T1]{fontenc}

\usepackage[utf8]{inputenc}

\usepackage{microtype}

\usepackage{inconsolata}

\usepackage{amssymb}
\usepackage{amsmath}
\usepackage{graphicx}
\usepackage{comment}
\usepackage{booktabs}
\usepackage{adjustbox}
\usepackage{multirow}
\usepackage{makecell}
\usepackage{algorithm}
\usepackage[noend]{algpseudocode}

\title{Continual Test-time Adaptation for End-to-end Speech Recognition \\on Noisy Speech}

\author{Guan-Ting Lin$^{\star}$, Wei-Ping Huang$^{\star}$, Hung-yi Lee \\
  Graduate Institute of Communication Engineering, National Taiwan University \\
  Taiwan \\
  \texttt{\{f10942104, hungyilee\}@ntu.edu.tw}, \texttt{thomas1232121@gmail.com} \\}

\begin{document}
\maketitle
\begin{abstract}
Deep Learning-based end-to-end Automatic Speech Recognition (ASR) has made significant strides but still struggles with performance on out-of-domain samples due to domain shifts in real-world scenarios. Test-Time Adaptation (TTA) methods address this issue by adapting models using test samples at inference time. However, current ASR TTA methods have largely focused on non-continual TTA, which limits cross-sample knowledge learning compared to continual TTA. In this work, we first propose a Fast-slow TTA framework for ASR that leverages the advantage of continual and non-continual TTA. Following this framework, we introduce Dynamic SUTA (\textbf{DSUTA}), an entropy-minimization-based continual TTA method for ASR. To enhance DSUTA robustness for time-varying multi-domain data, we design a \textbf{dynamic reset strategy} to automatically detect domain shifts and reset the model. Our method demonstrates superior performance on various noisy ASR datasets, outperforming both non-continual and continual TTA baselines while maintaining robustness to domain changes without requiring domain boundary information\footnote{The source code is available at \href{https://github.com/hhhaaahhhaa/Dynamic-SUTA}{https://github.com/\\hhhaaahhhaa/Dynamic-SUTA}}. 
\end{abstract}

\section{Introduction}

Deep learning-based end-to-end Automatic Speech Recognition (ASR) has made remarkable progress in recent years, achieving low recognition error rates for in-domain samples. However, domain shifts frequently occur in real-world scenarios. Although recent large-scale ASR models exhibit some generalization to out-of-domain test samples, their performance on out-of-domain samples still lags behind the in-domain performance.

Test-time adaptation (TTA) is an attractive method to address domain shift issues during inference time. TTA adapts the model using only single or batched test samples without needing the source training data at testing time. Specifically, the source model is adapted via unsupervised objectives like Entropy Minimization (EM) ~\citep{tent} or Pseudo-Labeling (PL)~\citep{pl} in inference time. 
TTA methods can be characterized into two categories: 
\textbf{1) Non-continual TTA} methods adapt the source model for each test utterance and reset to the original model for subsequent samples~\citep{tent}, and \textbf{2) Continual TTA} (CTTA) continuously adapts the model for target domains, leveraging knowledge learned across samples to improve performance~\citep{eata, stable-tta, rdumb}.

\begin{figure}[t]{}
\centering\includegraphics[width=1\linewidth]{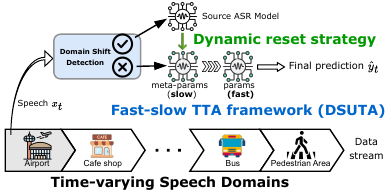}
    \caption{Illustration of the proposed Fast-slow TTA framework and dynamic reset strategy with time-varying speech domains. The Fast-Slow TTA framework includes meta-parameters that update \textit{slowly} to capture cross-domain knowledge, while other parameters update \textit{fast} for the incoming test samples. 
    The \textbf{Dynamic reset strategy} automatically detects domain shifts and resets the model to the source model.}
    \label{fig:overview}
\end{figure}

TTA methods initially strive in the field of computer vision~\citep{tent, eata, stable-tta, rdumb}. 
In speech recognition, recent studies have tailored TTA methods with EM-based optimization~\citep{suta, sgem, advancing}, proposing new training objectives and demonstrating effectiveness across datasets. However, existing ASR TTA methods only 
 focus on non-continual TTA, constraining the model to learn knowledge across samples. There is limited research on CTTA for end-to-end ASR. Recently, AWMC~\citep{awmc} proposed a pseudo-labeling CTTA method for ASR on a single test domain. However, as shown in previous work~\citep{suta}, pseudo-labeling is not as effective as EM-based methods, and its ability on long multi-domain testing data is unknown.

In this work, we propose a general \textbf{Fast-slow TTA} framework that leverages the advantages of both continual and non-continual TTA. Based on this framework, we introduce an EM-based CTTA method named \textbf{D}ynamic \textbf{SUTA} (\textbf{DSUTA}) for ASR. Furthermore, to enhance the robustness of DSUTA on time-varying domain data, we propose a \textbf{dynamic reset strategy} to automatically detect domain shifts and determine when to reset the model to the original source model. This strategy improves Fast-slow TTA over long sequences of multi-domain data streams.

We demonstrate the effectiveness of our method on single-domain and multi-domain time-varying ASR benchmarks under different acoustic conditions, simulating real-world changing environments. Our method outperforms the strong single-utterance baseline SUTA~\citep{suta} and the CTTA baseline AWMC~\citep{awmc}, showing robustness to domain changes even without knowing the domain boundaries.

Our contributions can be summarized as follows: 
\begin{enumerate}
    \item Propose the Fast-slow TTA framework to bridge the gap between continual and non-continual TTA.
    \item Introduce a specific version of the Fast-slow TTA method named \textbf{DSUTA} with a novel dynamic reset strategy to stabilize CTTA over multi-domain and long test data streams.
    \item Demonstrate significant improvement over both non-continual and continual baselines on single-domain and time-varying data.
\end{enumerate}

\section{Related Works}

\subsection{Non-continual TTA for ASR}
Non-continual TTA methods adapt the source model for each test utterance and reset to the original model for subsequent samples. SUTA~\citep{suta} introduces the first TTA approach for non-autoregressive ASR, based on entropy minimization and minimum class confusion. SGEM~\citep{sgem} extends TTA to autoregressive ASR models by introducing a general form of entropy minimization. ~\citet{advancing} enhances TTA with confidence-enhanced entropy minimization and short-term consistency regularization. However, these non-continual TTA methods view each utterance independently, which only relies on a single utterance and fails to leverage the knowledge across a stream of test samples to improve the adaptation.

\subsection{Continual TTA} 
Unlike non-continual TTA, which resets to the source model for each sample, continual TTA enables the online model to use learned knowledge to handle gradual changes in the target domain. However, it may suffer from model collapse if adaptation is unstable when the data stream is too long. To improve the performance and stability of CTTA, studies in the computer vision field have developed solutions like stochastic model restoring~\citep{cotta}, sample-efficiency entropy minimization~\citep{eata}, sharpness-aware reliable entropy minimization~\citep{stable-tta}, and fixed frequency model reset~\citep{rdumb}.

In the ASR research, \textit{there are limited studies on CTTA ASR}. Recently, AWMC~\citep{awmc} attempts continual TTA on ASR using a pseudo-labeling approach with an extra anchor model to prevent model collapse. However, AWMC~\citep{awmc} only measures the performance on single-domain data with the pseudo-labeling method. This work focuses on multi-domain time-varying long data streams. We propose a fast-slow TTA framework and dynamic reset strategy based on an entropy minimization-based CTTA method, which achieves better performance and stability.

\section{Methodology}
Section~\ref{sec:bridge} describes the proposed \textbf{Fast-slow TTA framework}. Following this framework, Section~\ref{sec:DSUTA} extends SUTA into \textbf{Dynamic SUTA}. To handle multi-domain scenarios better, we propose a \textbf{dynamic reset strategy} in Section~\ref{sec:DSUTA+reset}.

\subsection{Fast-slow TTA Framework}
\label{sec:bridge}
Non-continual TTA treats each sample as an independent learning event. The adaptation process can fit the current sample without affecting future samples; however, the learned knowledge cannot be transferred to future samples. 
In contrast, continual TTA utilizes learned knowledge, but overfitting the current sample can adversarially degrade performance on future samples.
For instance, if the model overly fits the current sample and (model collapse), the performance on future samples will significantly degrade with continual TTA, whereas in non-continual TTA, the performance remains unaffected.

\begin{figure}[t]{}
    \centering
\includegraphics[width=1\linewidth]{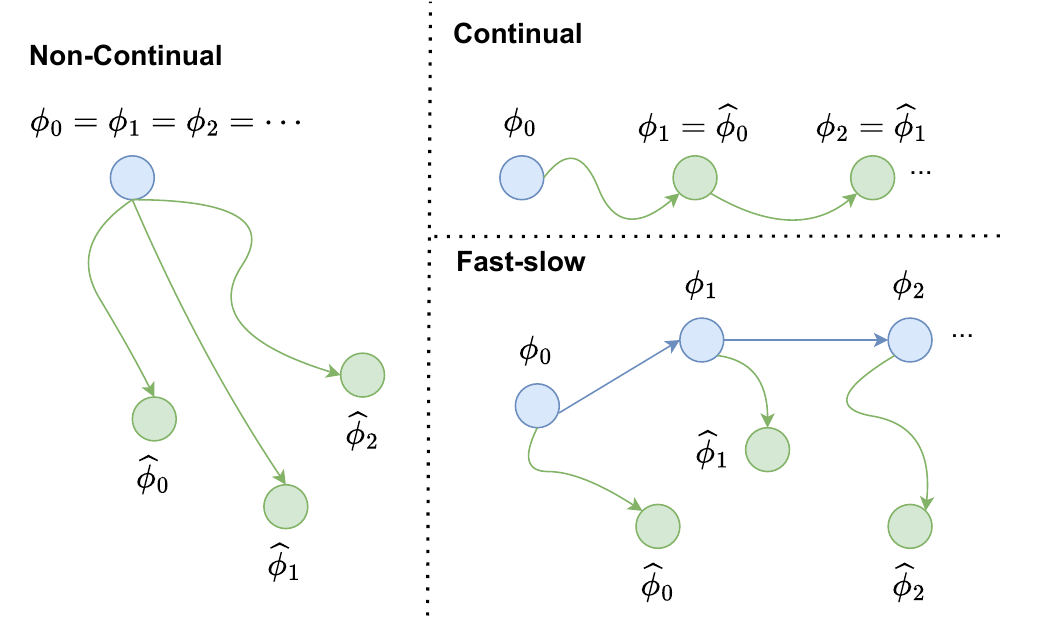}
    \caption{Illustration of the 3 different TTA approaches.}
    \label{fig:scheme}
\end{figure}

We propose \textbf{Fast-slow TTA}, a new CTTA framework that leverages learned knowledge while retaining the benefits of non-continual TTA, as shown in Figure~\ref{fig:scheme}. Fast-slow TTA aims to learn meta-parameters $\phi_t$ which evolve slowly over time. Instead of always starting the adaptation process from the pre-trained parameters, as in non-continual TTA, we start from $\phi_t$ at time step $t$. Specifically,
\begin{align*}
    \phi_0 &= \phi_{pre}, \\
    \widehat{\phi}_t &= A(\phi_t, x_t), \\
    \widehat{y}_t &= \widehat{\phi}_t(x_t), \\
    \phi_{t+1} &= U(\phi_{t}, x_t),
\end{align*}
where $\phi_{pre}$ are the pre-trained parameters, and $A$ and $U$ represent an adaptation algorithm and an update algorithm, respectively. The evaluation is based on the online predictions $\widehat{y}_t$.

\begin{algorithm}[]
\caption{Dynamic SUTA}\label{alg:dsuta}
\begin{algorithmic}[1]
\Require Data stream $\{x_t\}^{T}_{t=1}$, buffer $\mathcal{B}$ with size $M$, adaptation step $N$, pre-trained param $\phi_{pre}$
\Ensure Predictions $\{\widehat{y}_t\}^{T}_{t=1}$
\State $\mathcal{B}, \phi_1 \gets \{\}, \phi_{pre}$
\State $Results \gets \{\}$
\For{$t=1$ to $T$}
    \State $\widehat{\phi}_t \gets \phi_t$  \Comment{Adapt parameters}
    \For{$n=1$ to $N$}
        \State $\mathcal{L} \gets \mathcal{L}_{suta}(\widehat{\phi}_t, x)$
        \State $\widehat{\phi}_t \gets \textrm{Optimizer}(\widehat{\phi}_t, \mathcal{L})$
    \EndFor
    \State $\widehat{y}_t \gets \widehat{\phi}_t(x_t)$  \Comment{Save prediction}
    \State $Results \gets Results \cup \{\widehat{y}_t\}$

    \State $\mathcal{B} \gets \mathcal{B}\cup \{x_t\}$
    \If{$t \% M = 0$}\Comment{Update meta-param}
        \State $\mathcal{L} \gets \frac{1}{M}\sum_{x\in\mathcal{B}}\mathcal{L}_{suta}(\phi_{t}, x)$
        \State $\phi_{t+1} \gets \textrm{Optimizer}(\phi_{t}, \mathcal{L})$
        \State $\mathcal{B} \gets \{\}$
    \Else
        \State $\phi_{t+1} \gets \phi_t$
    \EndIf
\EndFor
\\
\Return $Results$
\end{algorithmic}
\end{algorithm}

The meta-parameters $\phi_t$ can leverage knowledge across samples. These parameters are slowly updated by $U$, and the final prediction is made after a fast adaptation $A$. This allows the parameters to fit the current sample for greater improvement while mitigating the risk of model collapse over time.

Fast-slow TTA generalizes to continual and non-continual TTA. If $U(\phi_t, x_t) = \phi_t$, i.e., $\phi_{t}$ remains constant over time, the framework degenerates to non-continual TTA. 
On the other hand, if $A=U$, i.e. $\phi_{t+1} = \widehat{\phi}_t$, the framework degenerates to continual TTA. 

%

\subsection{Dynamic SUTA}
\label{sec:DSUTA}
We propose \textbf{D}ynamic \textbf{SUTA} (DSUTA), a fast-slow TTA method based on SUTA~\citep{suta}. Specifically, given pre-trained parameters $\phi_{pre}$, for every incoming sample $x_t$, SUTA adapts $\phi_{pre}$ for $N$ steps with the objective $\mathcal{L}_{suta}$. $\mathcal{L}_{suta}$ consists of entropy loss and minimum class confusion loss. Entropy minimization aims to sharpen class distribution, and minimum class confusion aims to reduce the correlation between different prediction classes. See Appendix~\ref{sec:app-suta-loss} for the detailed loss function. Model parameters are reset to $\phi_{pre}$ when the next sample arrives.

For DSUTA, the adaptation algorithm $A$ is set exactly the same as SUTA, which iteratively adapts $\phi_t$ for $N$ steps with $\mathcal{L}_{suta}$ on $x_t$. To construct the update algorithm $U$, we introduce a small buffer $\mathcal{B}$ with size $M$. For every $M$ step, the buffer is filled and we calculate $\mathcal{L}_{suta}$ from these $M$ samples to update the meta-parameters $\phi_t$ with gradient descent. The buffer is then cleared. Thus, the meta-parameters $\phi_t$ gradually evolve by mini-batch gradient descent with batch size $M$. DSUTA can be viewed as a variant of SUTA, which starts the adaptation from dynamically changing $\phi_t$ instead of the fixed $\phi_{pre}$. Denote $\mathcal{L}_{suta}(\phi, x)$ as the loss of sample $x$ on model $\phi$. Algorithm~\ref{alg:dsuta} describes the pseudo code of DSUTA.

\subsection{DSUTA with Dynamic Reset Strategy}
\label{sec:DSUTA+reset}

\begin{figure*}[t]
    \centering
    \includegraphics[width=1\linewidth]{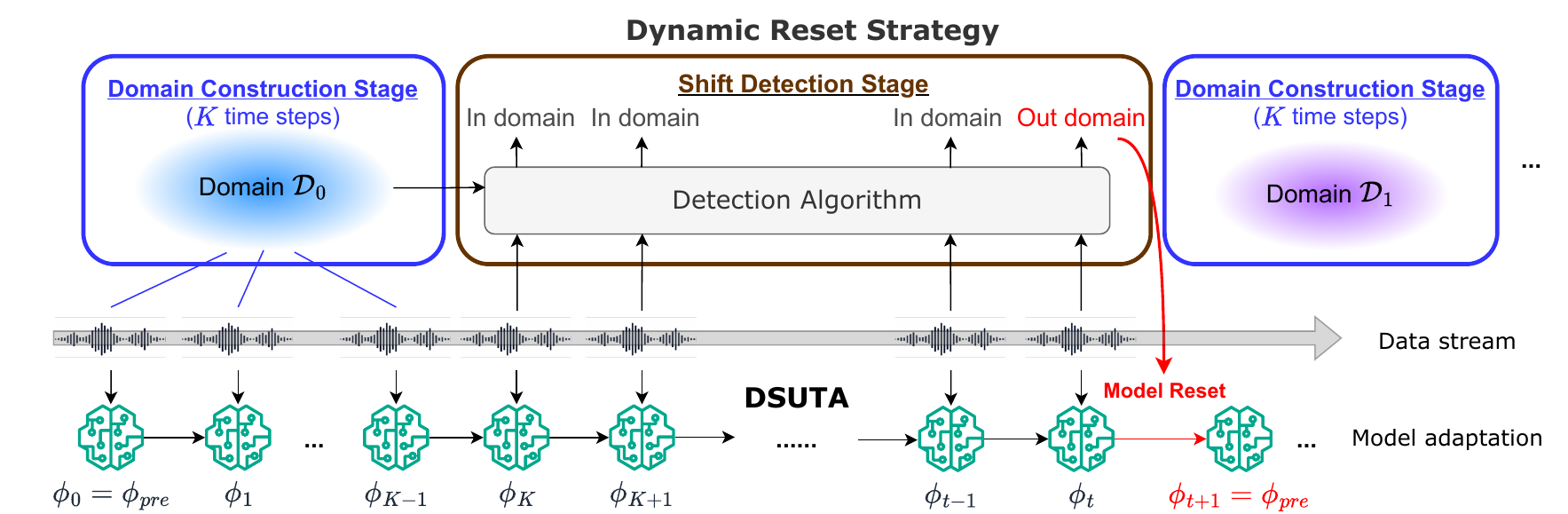}
    \caption{Sketch of DSUTA with the dynamic reset strategy. The domain construction stage and the shift detection stage alternate over time. When a large shift is detected, apply model reset to DSUTA, i.e., update $\phi_{t+1}=\phi_{pre}$.}
    \label{fig:pipeline}
\end{figure*}

As time progresses and the testing domain changes, multiple domain shifts significantly challenge the robustness of continual TTA methods. Recently, \citet{rdumb} has shown that \textit{model reset} at a fixed frequency, which resets the current parameters to the pre-trained ones at regular intervals, is a simple yet effective strategy. Therefore, we attempt to utilize \textit{model reset} strategy to update the meta-parameters $\phi_t$ in DSUTA\footnote{Non-continual TTA can be viewed as the case where we apply model reset at every time step.}. However, determining the optimal reset frequency in reality is challenging. To automatically determine when to apply model reset to $\phi_t$, we propose a \textbf{dynamic reset strategy} that actively detects large distribution shifts and dynamically resets $\phi_{t+1} = \phi_{pre}$.

Figure~\ref{fig:pipeline} provides an illustration of DSUTA with the dynamic reset strategy. Since distribution shift is a relative concept that is well-defined only after a base domain is constructed, we designed a \textbf{\textit{domain construction stage}} and a \textbf{\textit{shift detection stage}}. Our proposed method alternates between these two stages over time. The domain construction stage first constructs a base domain $\mathcal{D}$ with $K$ samples. No model reset will be applied during this stage. In the subsequent shift detection stage, a detection algorithm checks each incoming sample to determine if there is a significant distribution shift. If a large shift is detected, we apply model reset and switch to a new domain construction stage.

The following subsections describe the strategy in detail. We first introduce the Loss Improvement Index in Section \ref{sec:LII}, which measures the extent of the distribution shift. Then we define the domain construction stage and the shift detection stage in Section \ref{sec:dsuta+reset}.

\subsubsection{Loss Improvement Index (LII)}
\label{sec:LII}
We aim to find an indicator that measures the extent of the distribution shift from the base domain $\mathcal{D}$. To identify an appropriate indicator, we observed that given a model $\phi_\mathcal{D}$ trained on domain $\mathcal{D}$, $\mathcal{L}_{suta}$ for in-domain samples is empirically lower than that for out-of-domain samples. This suggests that $\mathcal{L}_{suta}(\phi_\mathcal{D}, x_t)$ might be a good indicator. Additionally, we found that subtracting the loss from the pre-trained model, $\mathcal{L}_{suta}(\phi_{pre}, x_t)$, is beneficial to normalize the inherent difficulty introduced by the data sample itself\footnote{See Section~\ref{sec:lii_design} for more discussion on indicator choice.}. Overall, we define \textbf{L}oss \textbf{I}mprovement \textbf{I}ndex (LII) as our indicator:
\[
    LII_t = \mathcal{L}(\phi_{\mathcal{D}}, x_t) - \mathcal{L}(\phi_{pre}, x_t),
\]
where $\mathcal{L}=\mathcal{L}_{suta}$. The construction of $\phi_{\mathcal{D}}$ will be described in the next section.

\subsubsection{Domain Construction Stage and Shift Detection Stage}
\label{sec:dsuta+reset}
We integrate DSUTA with the dynamic reset strategy as follows. Assume the model has been reset at time step $r$.\\
\textbf{(1) Domain Construction Stage}:
\begin{itemize}
    \item[1.] Let $k=\lfloor\frac{K}{2}\rfloor$, construct $\phi_{\mathcal{D}} = \phi_{r+k}$.
    \item[2.] Collect $LII_t$ for $t\in [r+k+1, r+K]$.
    \item[3.] At the end of the stage (i.e., $t=r+K$), compute $\mathcal{G}_{\mathcal{D}}=\mathcal{N}(\mu, \sigma^2)$ from the collected LIIs.
\end{itemize}
The goal is to estimate the distribution of LII. We construct $\phi_{\mathcal{D}}=\phi_{r+k}$ as the meta-parameters after observing $k$ samples since the last reset. Calculating the LII requires $\phi_{\mathcal{D}}$, and since TTA is an online process, $K-k$ is the number of LIIs we can collect for statistical estimation. A smaller $k$ might not suffice for $\phi_{\mathcal{D}}$ to adequately represent the domain, while a larger $k$ reduces the number of data points we can gather for estimation. Therefore, we empirically set $k=\lfloor\frac{K}{2}\rfloor$. \\
\textbf{(2) Shift Detection Stage}: 
\begin{equation*}
\phi_{t+1} = \begin{cases} 
    \phi_{pre},  & \textrm{if }\frac{LII_t-\mu}{\sigma} > 2, \\
    U_{DSUTA}(\phi_t, x_t), & \textrm{otherwise},
\end{cases}
\end{equation*}
where $U_{DSUTA}$ is the update algorithm of DSUTA.

During the domain construction stage, we develop a statistical model $\mathcal{G}_{\mathcal{D}}$ using $K-k$ samples to estimate the distribution of LII. In the shift detection stage, we trigger a reset operation if the LII exceeds a certain threshold, indicating an abnormally large shift. To determine whether the LII indicates such a shift, we conduct a right-tailed hypothesis test.

For the right-tailed hypothesis test, the common practice with a significance level of 0.05 corresponds to a Z-score of 1.64. Here, we use a Z-score of 2 for simplicity, which makes the condition for resetting slightly stricter.

Additionally, using the LII of a single sample for the hypothesis test is too sensitive. The \textbf{averaged LII} from multiple samples reduces variance and yields more reliable results. With DSUTA, we perform the hypothesis test every $M$ step, using the $M$ samples in DSUTA's buffer to calculate the averaged LII. The final shift detection stage is defined as follows:\\
\begin{equation*}
\phi_{t+1} = \begin{cases} 
    \phi_{pre}, \quad\textrm{if }\frac{1}{M}\sum_{i| x_i\in\mathcal{B}}\frac{LII_i-\mu}{\sigma / \sqrt{M}} > 2, M|t,\\
    U_{DSUTA}(\phi_t, x_t), \quad\textrm{otherwise}.
\end{cases}
\end{equation*}
Here, $\mathcal{B}$ represents the buffer containing the most recent $M$ samples. In our implementation, we further introduce a patience parameter $P$ to enhance the stability. Please refer to the Appendix Algorithm \ref{alg:dsuta-reset} for details.

\section{Experiments}
\subsection{Dataset}
\subsubsection{Single-domain Simulated Noisy Data}
\textbf{Corrupted Librispeech (LS-C)}: we follow previous works~\citep{sgem} by adding background noises from MS-SNSD~\citep{ms-snsd} into Librispeech test set~\citep{librispeech}. The noises include air conditioner (AC), airport announcement (AA), babble (BA), copy machine (CM), munching (MU), neighbors (NB), shutting door (SD), typing (TP), and vacuum cleaner (VC). We also apply Gaussian noise (GS) as in ~\citep{suta}, resulting in 10 different noises in total. The Signal-to-Noise Ratio (SNR) is set to 5 dB.\footnote{\citep{sgem} reported using 10dB noise but their source code and results show that they use 5 dB.} 

\subsubsection{Multi-domain Time-varying Data}
We create three time-changing multi-domain test data streams by concatenating different corruptions from LS-C.\\
\textbf{(a) MD-Easy}: Noises in MD-Easy are determined by the relatively \textit{well-performed} noises of the pre-trained model  (See Table~\ref{tab:main-exp}). Five background noises, in the order AC$\rightarrow$CM$\rightarrow$TP$\rightarrow$AA$\rightarrow$SD, were used, with 500 samples for each noise, making a total of 2500 samples. \\
\textbf{(b) MD-Hard}: Noises in MD-Hard are determined by the relatively \textit{poor-performed} noises of the pre-trained model (See Table~\ref{tab:main-exp}). Five background noises, in the order GS$\rightarrow$MU$\rightarrow$VC$\rightarrow$BA$\rightarrow$NB, were used, with 500 samples for each noise, making a total of 2500 samples. \\
\textbf{(c) MD-Long}: We first sample a background noise from the 10 available background noises, then sample a data sequence with this noise, with a \textit{random length ranging from 20 to 500}. We repeat this process until the total length reaches 10,000.

\subsubsection{Multi-domain Real Noisy Data}
\textbf{CHiME-3~\citep{chime}}: a noisy version of WSJ corpus mixed with real speech recorded in four noisy environments (Cafe, Bus, Street, Pedestrian Area). In this work, different types of noisy speech are randomly distributed in a sequence across time.

\subsection{Baselines}
\subsubsection{Non-continual TTA Baselines}
\textbf{1) SUTA~\citep{suta}} leverages unsupervised objectives (entropy minimization and minimum class confusion) to reduce uncertainty and minimize class correlations. Temperature smoothing is applied to flatten the output probability distributions, addressing issues with over-confident predictions. The adaptation process involves iteratively optimizing the objective of entropy minimization and minimal class correlation. \textbf{2) SGEM~\citep{sgem}} propose a general form of entropy minimization with negative sampling.

\subsubsection{Continual TTA Baselines} \textbf{3) CSUTA} is a straightforward continual version of SUTA without resetting parameters. \textbf{4) AWMC~\citep{awmc}} utilizes the anchor model to generate initial pseudo labels, the chaser model updates itself using these pseudo labels for self-training, and the leader model refines predictions through an exponential moving average. \\

\subsection{Implementation Details}
We use the wav2vec 2.0-base model fine-tuned on Librispeech 960 hours\footnote{https://huggingface.co/facebook/wav2vec2-base-960h} as the source ASR model. For SUTA, we follow the official implementation\footnote{https://github.com/DanielLin94144/Test-time-adaptation-ASR-SUTA}, where an additional reweighting trick is applied on the minimum class confusion loss. The default adaptation step of SUTA is $N=10$, as specified in the original paper. For SGEM, we follow the official  implementation\footnote{https://github.com/drumpt/SGEM}. For CSUTA, we set the adaptation step to $N=1$ since we found that any higher value would cause severe model collapse. We re-implemented AWMC with wav2vec 2.0, as there is no official code, and all hyperparameters follow the original paper. For the proposed DSUTA, the default buffer size is $M=5$, and the adaptation step is $N=10$. To reduce GPU memory usage, we exclude samples with raw lengths longer than 20 seconds in all experiments. This removes about 1\% of the data. 

For hyperparameter search, we investigate batch sizes (M=3, 5, 10) and domain construction steps (K=50, 100, 200), and find out that our method is robust across different setups. For more details, please see the Appendix \ref{appendix:hyper} section.

\subsection{Results}
\subsubsection{Single Domain}

\begin{table*}
  \centering
  \adjustbox{width=0.93\textwidth}{
  \begin{tabular}{lcccccccccc|c}
    \toprule
     \textbf{Method} & \textbf{AA} & \textbf{AC} & \textbf{BA} & \textbf{CM} & \textbf{GS} & \textbf{MU} & \textbf{NB} & \textbf{SD} & \textbf{TP} & \textbf{VC}  & \textbf{CHiME-3} \\
    \midrule
    Source model  & 40.6 & 27.7 & 66.9 & 49.7 & 75.6 & 51.4 & 120.1 & 19.4 & 25.8 & 49.7 & 30.0 \\ \midrule
    \textit{Non-continual} \\
    \textbf{SUTA}  & 30.6 & 17.4 & 53.7 & 38.7 & 54.5 & 39.0 & 112.3 & 15.0 & 17.4 & 39.3 & 23.3 \\
    \textbf{SGEM} & 30.9 & 17.8 & 54.5 & 39.2 & 56.3 & 39.2 & 113.0 & \textbf{14.9} & 17.5 & 40.3 & 23.5 \\
    \midrule
    \textit{Continual} \\
    \textbf{CSUTA}  & 39.8 & 22.6 & 63.4 & 53.4 & 58.4 & 54.7 & 68.1 & 23.2 & 23.0 & 50.9 & 27.6 \\
    \textbf{AWMC} &  31.6 & 18.0 & 61.6 & 37.7 & 48.5 & 36.2 & 131.9 & 17.0 & 18.0 & 36.1 &  22.4 \\
    \midrule
    \textit{Fast-slow} \\
    \textbf{DSUTA} &   \textbf{25.9} & \textbf{15.4} & \textbf{33.2} & \textbf{33.5} & \textbf{37.0} & \textbf{28.4} & \textbf{36.3} & 15.5 & \textbf{15.6} & \textbf{29.9} & \textbf{21.7} \\
    \bottomrule
  \end{tabular}}
  \caption{WER (\%) of different TTA methods on LS-C with 10 types of noises and CHiME-3. Reported WER is averaged over 3 runs.}
  \label{tab:main-exp}
\end{table*}

We compare TTA performance on LS-C by Word Error Rate (WER) in Table~\ref{tab:main-exp}. DSUTA shows significant improvement compared to the baseline methods. It outperforms both non-continual and continual baseline methods by a large margin, except for the SD domain, where it still achieves a 15.5\% WER, close to SGEM's performance (14.9\%). Notably, on the NB domain, DSUTA achieves a 36.3\% WER compared to SUTA, which has a WER greater than 100\%, demonstrating the effectiveness of our method.

\begin{figure}[]{}
    \centering
\includegraphics[width=1\linewidth, page=4]{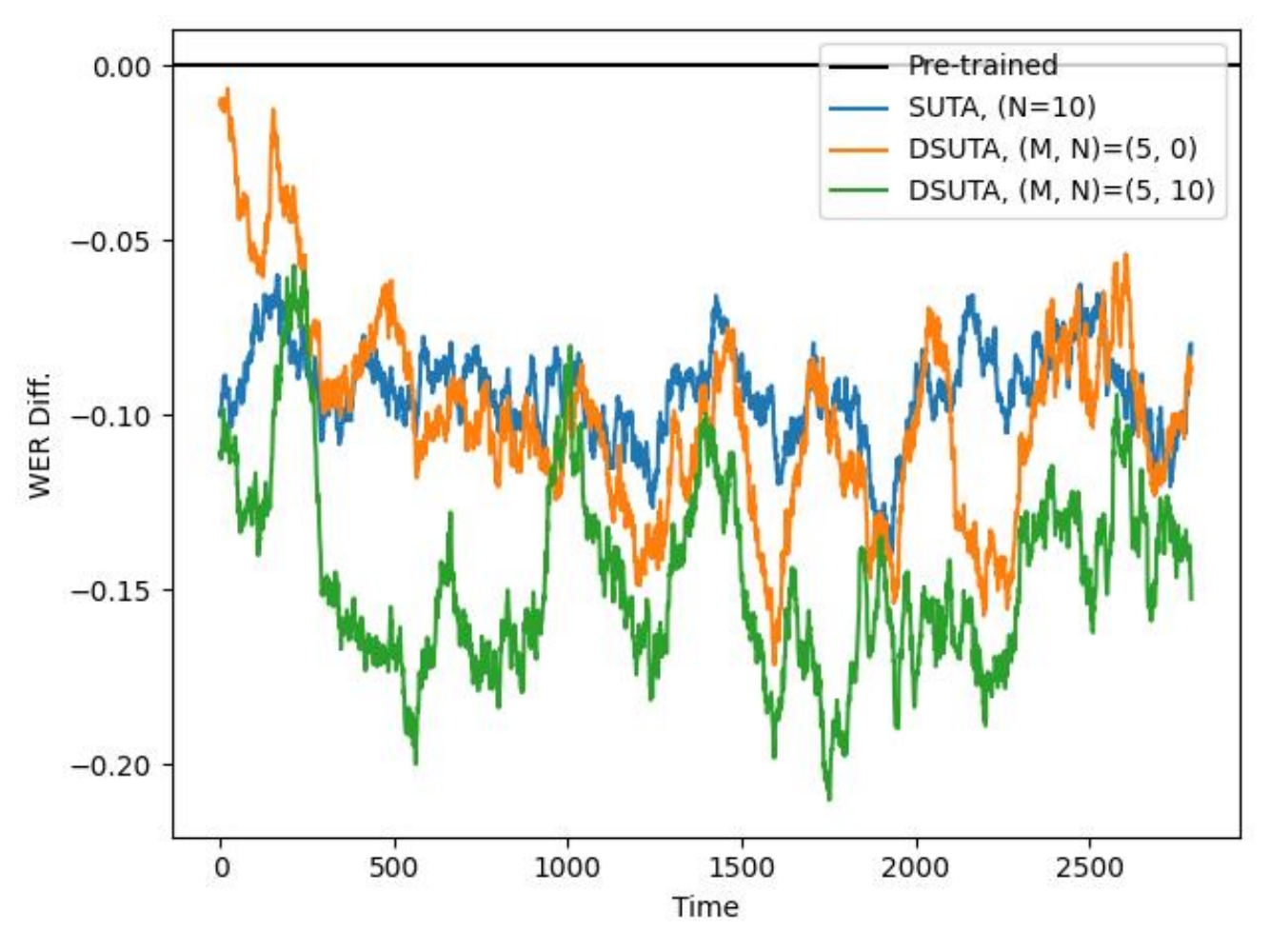}
    \caption{WER difference compared to the pre-trained model on CM domain over time. Data is smoothed by a window with a size of 100.}
    \label{fig:timeline-single}
\end{figure}

The key success factor of DSUTA is its ability to leverage learned knowledge from past samples. Figure~\ref{fig:timeline-single} plots the WER difference compared to the pre-trained model on the CM domain over time. We compare three methods: SUTA with $N=10$, DSUTA with $(M, N)=(5, 10)$, and DSUTA with $(M, N)=(5, 0)$, i.e., the learned $\phi_t$ itself. The WER of $\phi_t$ is lower than that of the pre-trained model, and DSUTA with 10-step adaptation outperforms SUTA with 10-step adaptation. In other words, DSUTA adaptation has a ``better start" compared to non-continual TTA methods due to the learned knowledge, resulting in superior performance. 


Table~\ref{tab:main-exp} also compares other continual TTA methods. Naive continual training, such as CSUTA, results in unsatisfactory performance and is sometimes even worse than the original pre-trained model due to its instability. Although AWMC is designed to increase stability, its performance sometimes lags behind SUTA, particularly in cases where the original pre-trained model has an extremely high error rate (BA and NB). This is not surprising since AWMC relies on a pseudo-label approach. In contrast, DSUTA uses mini-batch gradient descent to enhance stability without the use of pseudo labels. Furthermore, the fast-slow approach allows DSUTA to inherit SUTA's ability to better fit a single utterance, improving overall performance while avoiding the meta-parameters overfitting.

\subsubsection{Time-varying Multiple Domains}
\label{sec:multi-domain}
In the following experiment, we set DSUTA with $(M, N) = (5, 5)$ and compare DSUTA with dynamic reset strategy where $(M, N, K, P) = (5, 5, 100, 2)$ on multi-domain time-varying data. We also experiment DSUTA with two baseline reset strategies. 1) \textbf{\textit{Oracle boundary}} resets the model at the ground truth domain boundary, and 2) \textbf{\textit{Fixed reset}} is the simple fixed-frequency reset strategy, where the reset frequency is set to 50. 

\begin{table}
  \centering
  \adjustbox{width=\linewidth}{
      \begin{tabular}{lccc}
        \toprule
        \textbf{Method} & \textbf{MD-Easy} & \textbf{MD-Hard} & \textbf{MD-Long} \\
        \midrule
        Source model & 32.7 & 74.6 & 61.0 \\ \midrule
        \textit{Non-continual} \\
        \textbf{SUTA} & 24.0 & 60.4 & 53.3 \\
        \textbf{SGEM} & 25.0 & 61.0 & 53.4 \\
        \midrule
        \textit{Continual} \\
        \textbf{CSUTA} & 37.3 & 83.6 & 100.3 \\
        \textbf{AWMC} & 25.8 & 66.1 & 60.6 \\
        \midrule
        \textit{Fast-slow} \\
        \textbf{DSUTA} & 24.0 & 45.6 & 43.2 \\
        \textit{  w/ Dynamic reset} & \textbf{22.7} & \textbf{39.8} & \textbf{35.8} \\
        \textit{  w/ Fixed reset} & 22.8 & 49.4 & 45.2\\
        \textit{  \gray{w/ Oracle boundary}} & \gray{21.7} & \gray{36.9} & \gray{39.5}\\
        \bottomrule
      \end{tabular}
  }
  \caption{WER (\%) of different TTA methods on multi-domain time-varying data. Reported WER is averaged over 3 runs.}
  \label{tab:multi-exp}
\end{table}

Table~\ref{tab:multi-exp} summarizes the results. DSUTA is comparable to or better than other baseline methods, and applying \textit{Dynamic reset} further boosts the performance. Since we set DSUTA with fewer adaptation steps, our proposed method is both better and faster than SUTA in the multi-domain scenario.

For the non-continual TTA baselines, WER is improved in all cases but remains very high on MD-Hard and MD-Long. For the continual TTA baselines, CSUTA performs worse than the pre-trained model due to its instability. For AWMC, the original paper does not test in the multi-domain scenario, and our results show that AWMC is inferior to SUTA in this context.

Regarding the model reset strategy, the proposed \textit{Dynamic reset} outperforms \textit{Fixed reset}. \textit{Fixed reset} performs worse than DSUTA without reset on MD-Hard and MD-Long, suggesting that resetting too frequently might hinder the model from utilizing knowledge from past samples, thereby harming overall performance. Compared to \textit{Oracle boundary} (upper bound), \textit{Dynamic reset} achieves slightly worse performance on MD-Easy and MD-Hard. However, on MD-Long, \textit{Dynamic reset} surprisingly achieves a 35.8\% WER, which is even better than the 39.5\% WER using \textit{Oracle boundary}. Since \textit{Dynamic reset} automatically determines when to reset, it can further utilize the knowledge from other noise domains when it is beneficial,
rather than relying solely on single-domain data for adaptation.

Lastly, DSUTA demonstrates \textit{\textbf{superior performance on real multi-domain noisy data}}, as shown in Table \ref{tab:main-exp} column ``CHiME-3". DSUTA achieves 21.7\% WER, while the baseline SUTA and AWMC only yield 23.3\% and 22.4\% WER, respectively. This result further validates the proposed DSUTA can be generalized to real multi-domain noisy speech data stream. 

\section{Discussion}
\subsection{Why Choosing Averaged LII as an Indicator?}
\label{sec:lii_design}
\begin{figure}[t]{}
    \centering
\includegraphics[width=0.85\linewidth]{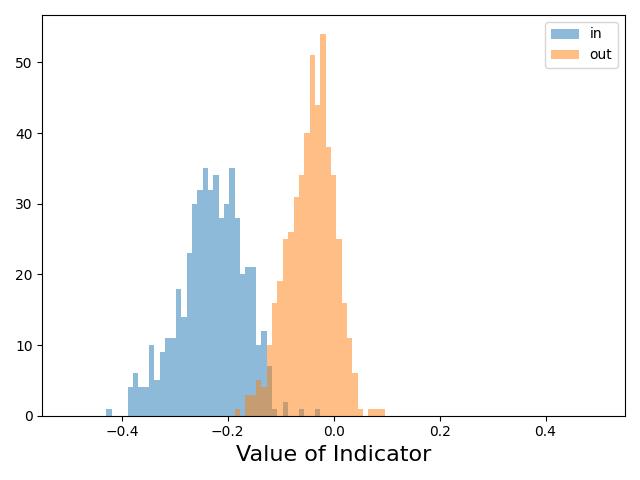}
    \caption{Distributions of averaged LII (over 5 samples) from the GS domain (in) and non-GS domains (out).}    \label{fig:lii}
\end{figure}

A good indicator should \textit{separate in-domain and out-of-domain samples into two clusters}. To visualize the indicator, we selected 500 samples from the GS domain as the source domain and randomly sampled 2000 samples from other domains as out-of-domain samples. $\phi_\mathcal{D}$ is then trained on 100 samples from the GS domain using $\mathcal{L}_{suta}$. We randomly sampled 500 averaged LIIs. Figure~\ref{fig:lii} visualizes the distributions of averaged LIIs (over 5 samples) of the remaining in-domain and out-of-domain samples. By using the averaged LII, two distributions are well separated.

\begin{figure}[t]{}
    \centering
\includegraphics[width=1\linewidth]{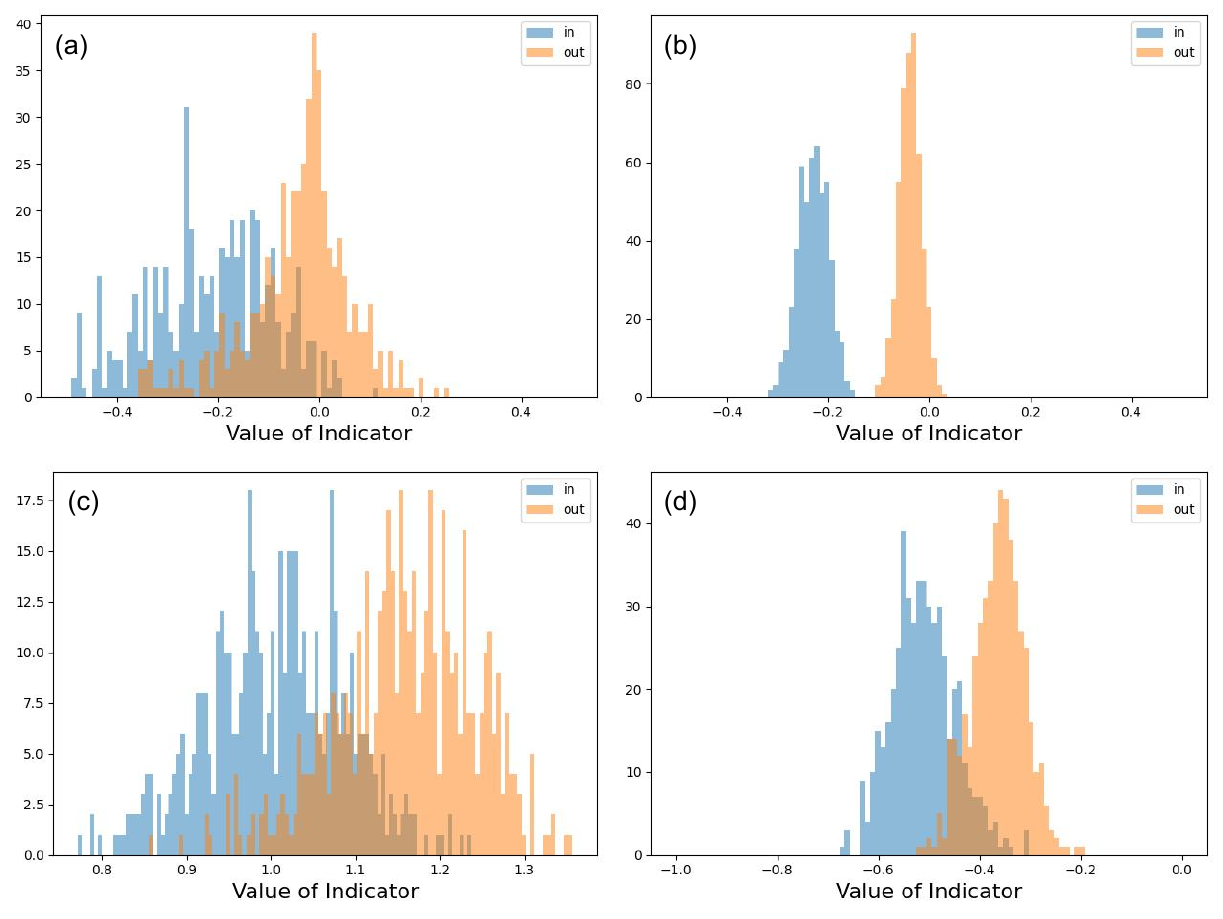}
    \caption{Distributions of other possible indicators. (a): original LII, (b): averaged LII for 20 samples, (c): without subtraction of pre-trained model loss, and (d): with the adapted parameters.}
    \label{fig:failed}
\end{figure}

Figure~\ref{fig:failed} visualizes the distributions of other possible choices of the indicator. Figure~\ref{fig:failed}a, b shows the distribution of averaged LII over 1 sample (i.e., the original LII) and 20 samples, respectively. Using a single sample is not sufficient to distinguish the distributions while considering more samples makes the detection more accurate. Figure~\ref{fig:failed}c illustrates the case without subtracting the loss from the pre-trained model, namely $\mathcal{L}(\phi_{\mathcal{D}}, x_t)$. The distributions are not well separated. In Figure~\ref{fig:failed}d, we also tried using the parameters after adaptation $A$ instead of the meta-parameters, namely
\[
    \mathcal{L}(A({\phi}_{\mathcal{D}}, x_t), x_t) - \mathcal{L}(A(\phi_{pre}, x_t), x_t).
\]
However, it resulted in more overlap between the two distributions than the proposed method.

\subsection{Different Domain Transition Rates}
In this section, we investigate \textit{how different domain transition rates affect the performance of reset strategies}. The original transition rate ($s$) of MD-Easy and MD-Hard is 500. We compare different reset strategies in 3 transition rates: $s=20, 100, 500$. To maintain a total length of the data stream to 2500, for $s=100$, the domain order sequence is repeated 5 times, and for $s=20$, the domain order sequence is repeated 25 times. We follow the hyperparameter settings described in Section~\ref{sec:multi-domain}.

\begin{table}
  \centering
  \adjustbox{width=\linewidth}{
    \begin{tabular}{lccc}
        \toprule
        \textbf{MD-Easy} & $s=20$ & $s=100$ & $s=500$ \\
        \midrule
        \textbf{DSUTA} & 24.1 & 23.9 & 24.0\\
        \textit{  w/ dynamic reset} & \textbf{23.8} & 23.7 & \textbf{22.7}\\
        \textit{  w/ fixed reset} & 24.6 & \textbf{23.1} & 22.8\\
        \gray{\textit{  w/ oracle boundary}} & \gray{23.7} & \gray{22.8} & \gray{21.7}\\
        \midrule
        \midrule
        \textbf{MD-Hard} & $s=20$ & $s=100$ & $s=500$ \\
        \midrule
        \textbf{DSUTA} & 45.6 & 44.7 & 45.6\\
        \textit{  w/ dynamic reset} & \textbf{42.3} & \textbf{44.5} & \textbf{39.8}\\
        \textit{  w/ fixed reset} & 53.3 & 49.9 & 49.4\\
        \gray{\textit{  w/ oracle boundary}} & \gray{57.3} & \gray{46.6} & \gray{36.9}\\
        \bottomrule
      \end{tabular}
  }
  \caption{WER (\%) of different reset strategies on MD-Easy and MD-Hard with different transition rates. Reported WER is averaged over 3 runs. $s$ is the domain transition rate.}
  \label{tab:transition-exp}
\end{table}

 The results are presented in Table~\ref{tab:transition-exp}. \textit{Oracle Boundary} and \textit{Fixed Reset} show that as the transition rate increases, resetting too often deteriorates performance. This phenomenon is more pronounced in MD-Hard, where DSUTA outperforms SUTA by a large margin, suggesting that continual learning is more effective in this context. \textit{Oracle Boundary} severely deteriorates performance when $s=20$ and $s=100$, implying that learning from samples from other noise domains might be beneficial. Since \textit{Dynamic Reset} automatically handles when to reset, it can utilize the knowledge from other noise domains, and reset is not triggered as frequently as in \textit{Oracle Boundary} or \textit{Fixed reset} under fast transitions, leading to better results.

In summary, the proposed \textit{Dynamic reset} offers good performance across diverse scenarios due to its flexibility. \textit{Dynamic reset} minimizes unnecessary resets and utilizes learned knowledge more effectively, consistently outperforming other reset strategies, making it a versatile solution.

\begin{table}[t]
  \centering
  \adjustbox{width=1\linewidth}{
      \begin{tabular}{lcccc}
        \toprule
          \multirow{2}{*}{\textbf{Method}} & \multicolumn{2}{c}{\textbf{Steps}} & \multicolumn{2}{c}{\textbf{Runtime (s)}} \\
         \cmidrule(lr){2-3} \cmidrule(lr){4-5}
         & \textbf{\#Forward} & \textbf{\#Backward} & \textbf{Total} & \textbf{Avg} \\
        \midrule
        \textit{Non-continual} \\
        \textbf{SUTA} & 100000 & 100000 & 5040 & 0.080\\
        \textbf{SGEM} & 100000 & 100000 & 11620 & 0.186\\
        \midrule
        \textit{Continual} \\
        \textbf{AWMC} & 300000 & 100000 & 11704 & 0.187\\
        \midrule
        \textit{Fast-slow} \\
        \textbf{DSUTA} & 52000 & 52000 & 3885 & 0.062\\
        \textbf{\textit{  w/ Dynamic reset}} & 72000 & 52000 & 4149 & 0.066\\
        \bottomrule
      \end{tabular}}
    \caption{Comparison of Forward/Backward steps and Runtime for different TTA methods on MD-Long. \textbf{Avg} is the averaged runtime (s) for a 1-second utterance. The result is averaged over 3 runs.}
  \label{tab:efficiency}
\end{table}

\subsection{Efficiency of the Proposed Method}
\label{sec:app-efficiency}
DSUTA is more efficient in adaptation steps than SUTA. Appendix Figure~\ref{fig:step-single} compares SUTA and DSUTA on 10 domains of LS-C under different adaptation steps $N=0, 1, 3, 5, 10$. DSUTA can use fewer adaptation steps to achieve better performance than SUTA with more adaptation steps.

To assess the efficiency of different TTA methods, we run them on MD-Long and compare the required forward/backward steps and runtime in Table~\ref{tab:efficiency}. CSUTA is excluded due to its poor performance. We follow the hyperparameter settings described in Section~\ref{sec:multi-domain}. All experiments were conducted on an Nvidia GeForce RTX 3080Ti GPU. Note that the results are for reference only, as values can slightly differ depending on the implementation. DSUTA is more efficient in the adaptation step and overall faster than SUTA, SGEM, and AWMC. Although adding the dynamic reset strategy slightly increases runtime, it remains faster overall. In conclusion, our method is not only superior in performance but also more efficient than existing approaches.

\subsection{Resets Frequency and Occurrence}
We propose a dynamic model reset strategy to detect domain shifts, improving both performance and efficiency. However, the frequency of model resets and their positions within the data stream remain unclear. In Table \ref{tab:reset_time_simple}, we present the reset timings of our method across three runs. The oracle boundaries occur at steps 500, 1000, 1500, and 2000. The results indicate that the reset timings are close to, but not exactly aligned with, the oracle boundaries.

\begin{table}[t]
  \centering
  \adjustbox{width=0.82\linewidth}{
      \begin{tabular}{lc}
        \toprule
        \textbf{3 runs} & \textbf{Automatic reset step} \\
        \midrule
        \textbf{MD-Easy-1} & 540, 1530, 2010 \\
        \textbf{MD-Easy-2} & 565, 1635, 2025 \\
        \textbf{MD-Easy-3} & 560, 1530, 2010 \\
        \midrule
        \textbf{MD-Hard-1} & 155, 555, 790, 1045 \\
        \textbf{MD-Hard-2} & 510, 1510, 2010 \\
        \textbf{MD-Hard-3} & 155, 510, 1165, 1510, 2010 \\
        \bottomrule
      \end{tabular}}
    \caption{Reset times and Automatic reset steps for MD-Easy and MD-Hard tasks over 3 runs. The ground truth task boundaries at steps equal to 500, 1000, 1500, and 2000. }
  \label{tab:reset_time_simple}
\end{table}

\section{Conclusion}


In this work, we advance the non-continual Test-Time Adaptation (TTA) method for ASR into a continual learning framework using a novel approach to stabilize adaptation and improve performance. Specifically, we introduce Dynamic SUTA (DSUTA), a fast-slow method that combines non-continual and continual TTA, demonstrating significant improvements on single-domain test data. Additionally, we propose a statistical dynamic reset strategy to enhance robustness and performance on time-varying test data streams. Experimental results indicate that our proposed method outperforms the non-continual SUTA baseline and previous continual TTA methods using pseudo labeling.

\section*{Limitations}
The primary limitations of this paper are as follows: 
\textbf{Domain Shift with Background Noises}:
In this work, we use noise corruptions to simulate changing domains and control domain shifts. However, there are various other speech domains to study, such as accents, speaker characteristics, and speaking styles. We will consider these domains in future research.
\\
\textbf{Different Types of End-to-End ASR Models}:
This work follows SUTA with a CTC-based ASR model, but there are different kinds of end-to-end ASR models available. As shown in \citep{sgem}, entropy minimization-based TTA methods can be extended to other end-to-end ASR models. We encourage future research to extend our DSUTA method to these other end-to-end ASR models. \\
\textbf{Not Addressing Model Forgetting}: This work focuses on adaptation to testing samples during inference time, rather than memorizing all past knowledge. Consequently, the proposed method might experience catastrophic forgetting as the domain changes. However, given a new test sample, the method can instantly adapt to that instance, ensuring that the final performance remains strong.

\section*{Acknowledgments}
The authors want to thank the reviewers for their insightful comments and suggestions during the rebuttal period. Guan-Ting Lin is supported by the NTU GICE and the NTU Ph.D. scholarship. 

\bibliography{custom}
\appendix

\section{Appendix}
\label{sec:appendix}

\subsection{Different noise levels}
\label{sec:noise_level}
From Table~\ref{tab:main-exp} and Table~\ref{tab:multi-exp}, we observe a trend that DSUTA has a larger advantage over other methods under severe domain shift where the pre-trained model performs poorly. To investigate \textit{how different levels of domain shift affect the proposed method}, we compare the pre-trained model, SUTA, and DSUTA with noise levels of 0dB, 5dB, and 10dB on the AC, SD, and TP domains from LS-C, which are the top 3 well-performing domains for the pre-trained model. We set $N=5$ for both SUTA and DSUTA. Table~\ref{tab:noise-exp} summarizes the results.

The results show that DSUTA is more effective under severe corruption. As the noise level decreases, although DSUTA outperforms the pre-trained model, SUTA becomes better than DSUTA. We hypothesize that while DSUTA is quite effective in noisy speech, its performance gain over the non-continual version (SUTA) is limited to relatively clean speech. Improving DSUTA's performance over SUTA on clean speech remains an area for future work.

\begin{table}
  \centering
      \begin{tabular}{cccccc}
        \toprule
        \textbf{Domain} & \textbf{Method} & \textbf{0dB} & \textbf{5dB} & \textbf{10dB} \\
        \midrule
        \multirow{3}{*}{AC}& Pre-trained & 63.7 & 27.7 & 14.2 \\
        & SUTA & 39.5 & 17.4 & \textbf{10.6} \\
        & DSUTA & \textbf{27.6} & \textbf{16.0} & 11.5 \\
        \midrule
        \multirow{3}{*}{SD}& Pre-trained & 29.7 & 19.4 & 13.6 \\
        & SUTA & 23.6 & \textbf{15.0} & \textbf{10.8} \\
        & DSUTA & \textbf{22.4} & 15.5 & 12.0 \\
        \midrule
        \multirow{3}{*}{TP}& Pre-trained & 42.4 & 25.8 & 16.6 \\
        & SUTA & 28.8 & 17.4 & \textbf{12.1} \\
        & DSUTA & \textbf{22.4} & \textbf{16.3} & 12.4 \\
        \bottomrule
      \end{tabular}
      \caption{WER(\%) comparison for different noise levels. Reported WER is averaged over 3 runs.}
  \label{tab:noise-exp}
\end{table}

\begin{figure*}
    \centering
\includegraphics[width=0.8\linewidth]{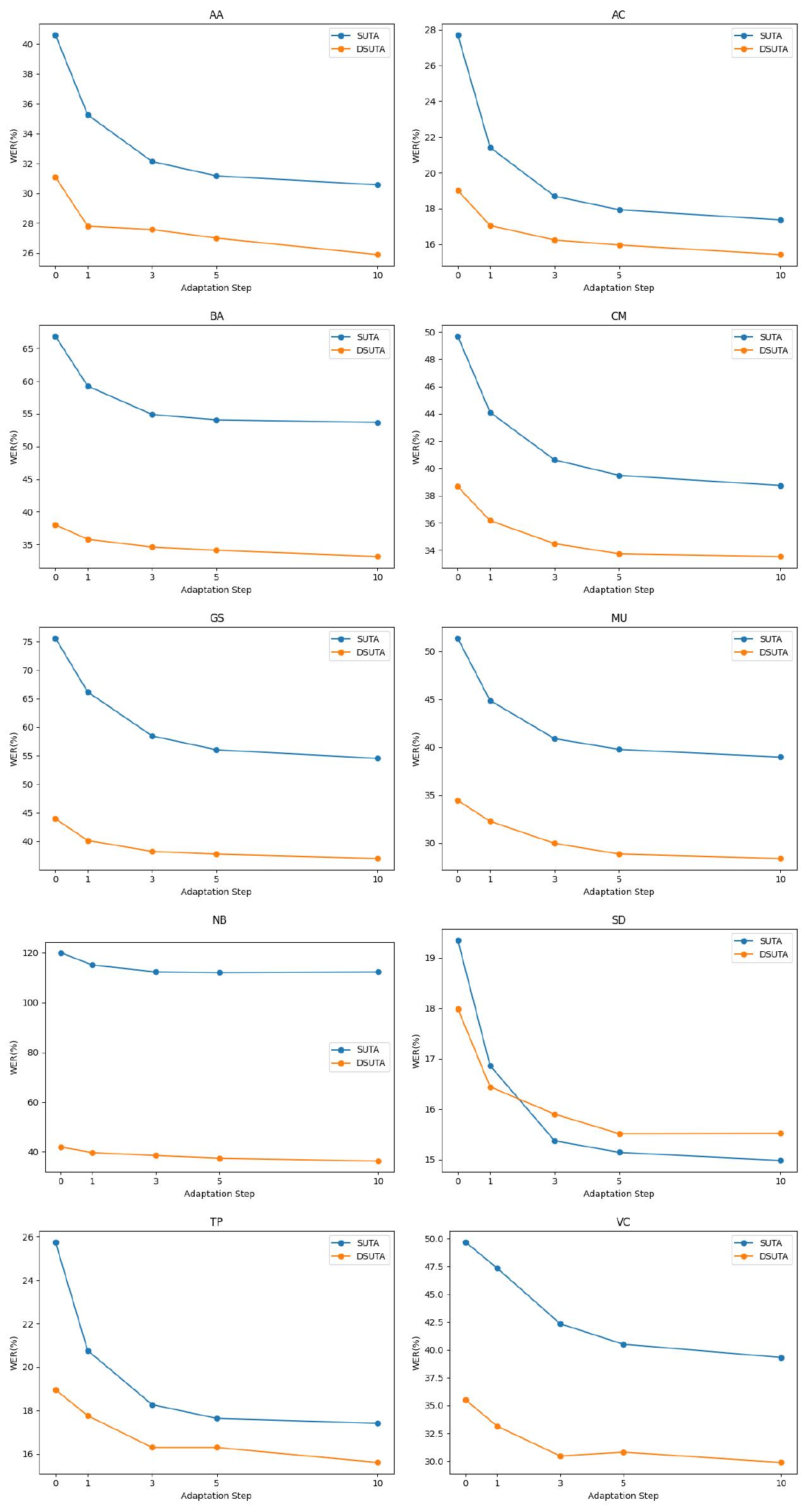}
    \caption{WER (\%) of different number of adaptation steps on 10 noise domains of LS-C.}
    \label{fig:step-single}
\end{figure*}

\subsection{Hyper-parameter Tuning}
\label{appendix:hyper}
We explore different hyper-parameters for DSUTA with the dynamic reset strategy. We use MD-Long as the data sequence. Table~\ref{tab:ablate} presents the results for various buffer sizes $M$. Our proposed method performs well overall. A smaller buffer size can make the update of meta-parameters unstable, while a larger buffer increases latency in triggering model reset after a domain shift since the shift is detected once every $M$ steps. Therefore, a medium buffer size is preferred.

Table~\ref{tab:ablate} also presents the results for different $K$ values during the domain construction stage. Again, our proposed method performs well overall. The performance of $K=50$ is worse than $K=100$ and $K=200$, suggesting that domain construction benefits from having enough steps to collect LII statistics and train a domain-specialized model $\phi_D$. 

\subsection{Generalization to Different Source ASR Models}

To test the generalization of the proposed method, we adopt other source ASR models with DSUTA and dynamic reset strategy. Table \ref{tab:ablate2} reports the results with the ASR model fine-tuned from wav2vec 2.0-base, data2vec-base\footnote{https://huggingface.co/facebook/data2vec-audio-base-960h}, and HuBERT-large\footnote{https://huggingface.co/facebook/hubert-large-ls960-ft} model. All the ASR models are trained with Librispeech 960 hours. Results show that both DSUTA and DSUTA with the dynamic reset strategy perform effectively across different models, yielding significantly better WER than the pre-trained model and the SUTA.

\begin{table}[t]
  \centering
    \begin{tabular}{cc}
        \toprule
        \textbf{Setup} & \textbf{WER} \\ \midrule
        $M=3$ & 36.8\\
        $M=5$ & \textbf{35.8}\\
        $M=10$ & 37.0\\
        \midrule
        $K=50$ & 38.5\\
        $K=100$ & 35.8\\
        $K=200$ & \textbf{35.5}\\
        \bottomrule
    \end{tabular}
  \caption{WER(\%) comparison of different hyperparameters on MD-Long. Reported WER is averaged over 3 runs.}
  \label{tab:ablate}
\end{table}

\begin{table}[t]
  \centering
  \adjustbox{width=\linewidth}{
    \begin{tabular}{lccc}
        \toprule
        \textbf{Method} & wav2vec2-base & data2vec-base & hubert-large\\ \midrule
        \textbf{Pre-trained} & 61.0 & 59.6 & 43.3\\
        \textbf{SUTA} & 53.3 & 53.3 & 39.3\\
        \textbf{DSUTA} & 43.2 & 52.0 & \textbf{17.8}\\
        \textbf{\textit{  w/ Dynamic reset}} & \textbf{35.8} & \textbf{46.3} & 19.0\\
        \bottomrule
    \end{tabular}
  }
  \caption{WER(\%) comparison of different CTC-based ASR models on MD-Long. Reported WER is averaged over 3 runs.}
  \label{tab:ablate2}
\end{table}

\label{sec:differet-asr}

\subsection{Objective of SUTA ($L_{suta}$)}
\label{sec:app-suta-loss}
Assume $C$ is the number of output classes and $L$ is the number of frames in the utterance. $\mathbf{P_{\cdot j}} \in \mathbb{R}^{L}$ denotes the output probabilities of the $j$-th class of the $L$ frames.\\
\textbf{Entropy Minimization (EM)}:
\begin{equation*}
    \mathcal{L}_{em} = \frac{1}{L} \sum_{i=1}^L \mathcal{H}_i = - \frac{1}{L} \sum_{i=1}^L \sum_{j=1}^C \mathbf{P_{ij}} \log \mathbf{P_{ij}}. 
\end{equation*}
\textbf{Minimum Class Confusion (MCC)}:
\begin{equation*}
    \mathcal{L}_{mcc} = \sum_{j=1}^C \sum_{j'\neq j}^C \mathbf{P_{\cdot j}^\top P_{\cdot j'}}. 
\end{equation*}\\

The final SUTA objective is defined as a mixture of $\mathcal{L}_{em}$ and $\mathcal{L}_{mcc}$:
\[
\mathcal{L}_{suta} =  \alpha \mathcal{L}_{em} + (1-\alpha) \mathcal{L}_{mcc}.
\]
We follow the settings in the original paper, which set $\alpha=0.3$ and apply temperature smoothing on logits with a temperature of $2.5$.

\begin{algorithm*}
\caption{Dynamic SUTA with the \color{red}dynamic reset strategy\color{black}}\label{alg:dsuta-reset}
\begin{algorithmic}[1]
\Require Data Sequence $\{x_t\}^{T}_{t=1}$, buffer $\mathcal{B}$ with size $M$, adaptation step $N$, \color{red}number of samples for construction $K$, patience $P$, \color{black}pre-trained parameters $\phi_{pre}$
\Ensure Predictions $\{\widehat{y}_t\}^{T}_{t=1}$
\State $\mathcal{B}, \phi_1 \gets \{\}, \phi_{pre}$
\color{red}
\State $k, last\_reset, stats \gets \lfloor K/2\rfloor, 0, \{\}$
\color{black}
\State $Results \gets \{\}$
\For{$t=1$ to $T$}
    \State $\widehat{\phi}_t \gets \phi_t$  \Comment{SUTA as adapt algorithm}
    \For{$n=1$ to $N$}
        \State $\mathcal{L} \gets \mathcal{L}_{suta}(\widehat{\phi}_t, x)$
        \State $\widehat{\phi}_t \gets \textrm{Optimizer}(\widehat{\phi}_t, \mathcal{L})$
    \EndFor
    \State $\widehat{y}_t \gets \widehat{\phi}_t(x_t)$  \Comment{Inference and save the prediction}
    \State $Results \gets Results \cup \{\widehat{y}_t\}$
    \State $\mathcal{B} \gets \mathcal{B}\cup \{x_t\}$
    \If{$t\% M= 0$}  \Comment{Update meta-parameter every $M$ steps}
        \color{red}
        \If{$t > last\_reset + K \textrm{ and IsReset}(\mathcal{G}, \mathcal{B}, P)$}  \Comment{Dynamic reset}
            \State $\phi_{t+1} \gets \phi_{pre}$
            \State $last\_reset \gets t$
        \Else
            \color{black}
            \State $\mathcal{L} \gets \frac{1}{M}\sum_{x\in\mathcal{B}}\mathcal{L}_{suta}(\phi_{t}, x)$
            \State $\phi_{t+1} \gets \textrm{Optimizer}(\phi_{t}, \mathcal{L})$
        \EndIf
        \State $\mathcal{B} \gets \{\}$
    \Else
        \State $\phi_{t+1} \gets \phi_t$
    \EndIf
    \color{red}
    \If{$t = last\_reset + k$}  \Comment{Save the domain-specialized model}
        \State $\phi_{\mathcal{D}} \gets \phi_t$
    \ElsIf{$last\_reset + k < t\leq last\_reset + K$}  \Comment{Collect LII stats}
        \State $stats \gets stats\cup\{LII_t\}$
    \EndIf
    \If{$t = last\_reset + K$}  \Comment{Generate distribution}
        \State $\mathcal{G} \gets \mathcal{N}(\mu_{stats}, \sigma^2_{stats})$
    \EndIf
    \color{black}
\EndFor\\
\Return $Results$
\end{algorithmic}
\end{algorithm*}

\end{document}